\documentclass[aps,nofootinbib,reprint,nobibnotes,showpacs,superscriptaddress]{revtex4-1}

\usepackage[english]{babel}
	\addto\captionsenglish{%
	}

\usepackage[babel]{csquotes}	

\usepackage{graphicx}

	\graphicspath{{graphics/}}

\usepackage{amsmath}
\usepackage{amssymb}
\usepackage[version=3]{mhchem}	
\usepackage{bm}	
\usepackage{nicefrac}	
\usepackage{mathrsfs}

\usepackage[hidelinks]{hyperref}
\hypersetup{
    colorlinks,
    linkcolor={red!70!black},
    citecolor={green!70!black},
    urlcolor={blue!70!black}
}

\usepackage[usenames,dvipsnames]{xcolor}	

\newcommand{\bb}{$0\nu \beta \beta$} 

\newcommand{\mbb}{m_{\beta \beta}} 

\newcommand{\NH}{$\mathcal{NH}$} 
\newcommand{\IH}{$\mathcal{IH}$} 

\newcommand{\meV}{\text{meV}}	\newcommand{\eV}{\text{eV}}

\newcommand{\el}{\text{e}}

\begin{document}

\title{The contribution of light Majorana neutrinos to neutrinoless double beta decay and cosmology}

	\author{Stefano Dell'Oro}
		\email{stefano.delloro@gssi.infn.it}
	\author{Simone Marcocci}
		\email{simone.marcocci@gssi.infn.it}
		\affiliation{INFN, Gran Sasso Science Institute, Viale F.\ Crispi 7, 67100 L'Aquila, Italy \\}%
	\author{Matteo Viel}
		\email{viel@oats.inaf.it}
		\affiliation{INAF, Osservatorio Astronomico di Trieste, Via G.\,B.\ Tiepolo 11, 34131 Trieste, Italy \\}%
		\affiliation{INFN, Sezione di Trieste, Via Valerio 2, 34127 Trieste, Italy \\}%
	\author{Francesco Vissani}
		\email{francesco.vissani@lngs.infn.it}
		\affiliation{INFN, Laboratori Nazionali del Gran Sasso, Via G.\ Acitelli 22, 67100 Assergi (AQ), Italy \\}%
		\affiliation{INFN, Gran Sasso Science Institute, Viale F.\ Crispi 7, 67100 L'Aquila, Italy \\}%

\date{\today}
	

	\begin{abstract}
		Cosmology is making impressive progress and it is producing stringent bounds on the sum of the neutrino masses 
		$\Sigma$, a parameter of great importance for the current laboratory experiments. 
		In this letter, we exploit the potential relevance of the analysis of Palanque-Delabrouille \emph{et al.} 
		[JCAP 1502, 045 (2015)] to the neutrinoless double beta decay (\bb) search.
		This analysis indicates small values for the lightest neutrino mass, since the authors find 
		$\Sigma<84\,\meV$ at 1$\sigma$\,C.\,L., and provides a $1\sigma$ preference for the normal hierarchy.
		The allowed values for the Majorana effective mass, probed by \bb, turn out to 
		be $< 75\,\meV$ at $3\sigma$\,C.\,L.\, and lower down to less than $20\,\meV$ at $1\sigma$\,C.\,L.\,.
		If this indication is confirmed, the impact on the \bb~experiments will be tremendous since the possibility of 
		detecting a signal will be out of the reach of the next generation of experiments. 
	\end{abstract}

	\pacs{%
		14.60.Pq, 
		23.40.-s, 
		90.80.Es 
		\hspace{97pt}%
		DOI: \href{http://dx.doi.org/10.1088/1475-7516/2015/12/023}{10.1088/1475-7516/2015/12/023}
	}

	\maketitle

\section{Introduction}

	Neutrinoless double beta decay (\bb) \cite{Furry:1939qr} probes lepton number conservation 
	and allows the investigation of the nature of the neutrino mass eigenstates.
	Assuming that ``ordinary'' light neutrinos mediate the \bb~transition, the key parameter that regulates 
	the decay rate is the Majorana Effective Mass, namely
	\begin{equation} \label{eq:mbb_def}
		\mbb \equiv \biggl| \el^{i\alpha_1}|U_{\el 1}^2|m_1 + \el^{i\alpha_2}|U_{\el 2}^2|m_2 + |U_{\el 3}^2|m_3 \biggr|.
	\end{equation}
	It represents the absolute value of the ee entry of the neutrino mass matrix. Here, $m_i$ are the masses 
	of the individual neutrinos $\nu_i$, $\alpha_{1,2}$ are the Majorana phases and 
	$U_{\el i}$ are the elements of the mixing matrix that defines the composition of
	the electron neutrino: $|\nu_\el \rangle~=~\sum_{i=1}^3 U_{\el i}^* |\nu_i \rangle$.
	From Eq.~(\ref{eq:mbb_def}), it is evident that the \bb~amplitude strongly depends upon the Majorana phase variation 
	and the absolute neutrino mass scale. 
	
	The mixing parameters and the squared mass differences are effectively measured in oscillation 
	experiments~\cite{Capozzi:2013csa}.
	Since $\alpha_1$ and $\alpha_2$ are very challenging to test~\cite{DellOro:2014yca}, we focus on the implications 
	coming from measurements of neutrino masses which, at present, are best probed by cosmological surveys.

\section{Results from cosmological data}

	The three light neutrino scenario is consistent with all known facts in particle physics including the new measurements 
	by Planck~\cite{Planck:2015xua}. In this assumption, the quantity $\Sigma$ is just the sum of the masses of 
	the neutrinos: 
	\begin{equation}
		\Sigma \equiv m_1 + m_2 + m_3.
	\end{equation} 
	The present information on three-flavor neutrino oscillations is compatible with two different neutrino mass spectra: 
	normal hierarchy (\NH) and inverted hierarchy (\IH). 
	In the former case one gets:
	\begin{equation} \label{eq:sigmaNH}
		\left\{
		\begin{aligned}
			m_1 &= m \\
			m_2 &= \sqrt{m^2 + \delta m^2} \\
			m_3 &= \sqrt{m^2 + \Delta m^2 + \delta m^2/2}
		\end{aligned}
		\right.
	\end{equation}
	while, in the latter:
	\begin{equation} \label{eq:sigmaIH}
		\left\{
		\begin{aligned}
			m_1 &= \sqrt{m^2 + \Delta m^2 - \delta m^2/2} \\
			m_2 &= \sqrt{m^2 + \Delta m^2 + \delta m^2/2} \\
			m_3 &= m.
		\end{aligned}
		\right.
	\end{equation}
	Here, $m$ is the lightest neutrino mass, while $\delta m^2$ and $\Delta m^2$ are the mass splittings measured by 
	oscillations, defined according to Ref.~\cite{Capozzi:2013csa}.
	
	A recent analysis~\cite{Gonzalez-Garcia:2014bfa}, based exclusively on oscillation data, finds a $1\sigma$ preference 
	for the \IH. This conclusion relies on partially contradictory information. 
	In fact, the analysis of atmospheric neutrinos performed by the Super-Kamiokande Collaboration~\cite{Wendell:2014dka} 
	shows a ``tendency'' (i.\,e. a $1\sigma$ preference) in favor of the \NH. 
	
	The determination of the mass hierarchy is a difficult challenge. In view of its great importance and of the 
	following discussion, we believe that future global analyses could benefit from the inclusion of the information 
	obtained from cosmology.

	In this work, we focus on the tightest experimental limits currently available for $\Sigma$.
	These are usually obtained by combining cosmic microwave background (CMB) data with the Lyman-$\alpha$ forest ones.
	Since they probe different length scales, the combination allows a more effective investigation of the neutrino 
	induced suppression in terms of matter power spectrum, both in scale and redshift.

	Among the different analyses performed, it is worth stressing the following $2\sigma$ C.\,L. upper limits:
	0.17\,eV obtained in Ref.\ \cite{Seljak:2006bg} by combining CMB data of the Wilkinson Microwave Anisotropy Probe, 
		galaxy clustering and the Lyman-$\alpha$ forest of the Sloan Digital Sky Survey (SDSS); 
	0.18\,eV of Ref.\ \cite{Riemer-Sorensen:2013jsa} using Planck and WiggleZ galaxy clustering data; 
	0.14\,eV obtained in Ref.\ \cite{Costanzi:2014tna} by combining Lyman-$\alpha$ SDSS data with Planck; 
	0.17\,eV obtained in Ref.\ \cite{Planck:2015xua} by using Planck temperature and polarization measurements including a 
		prior on the Hubble parameter, Supernovae and Baryonic Acoustic Oscillations (BAOs).
	
	More recently, by using a new sample of quasar spectra from SDSS-III and Baryon Oscillation Spectroscopic Survey searches 
	and a novel theoretical framework which incorporates neutrino non-linearities self consistently, Palanque-Delabrouille 
	\emph{et al.}~\cite{Palanque-Delabrouille:2014jca} have obtained a new tight limit on $\Sigma$.
	This constraint was derived both in frequentist and bayesian statistics by combining the Planck 2013 
	results~\cite{Ade:2013zuv} with the one-dimensional flux power spectrum measurement
	of the Lyman-$\alpha$ forest of Ref.\ \cite{Palanque-Delabrouille:2013gaa}.

	The one-dimensional flux power spectrum is the key observable used to derive the tight upper bound on $\Sigma$. 
	The final results do not depend upon cosmological and astrophysical parameters since the constraints obtained on the 
	total neutrino mass $\Sigma$ derived in Ref.~\cite{Palanque-Delabrouille:2014jca} include both statistical and 
	systematic uncertainties.
	The error budget is dominated by systematic uncertainties, that are treated as nuisance parameters affecting the
	observed flux power spectrum.
	
	A detailed modeling and an accurate description of the sources of uncertainties that affect this measurement is 
	quantitatively presented in Ref.~\cite{Palanque-Delabrouille:2013gaa} and we refer to this paper for a more 
	exhaustive discussion (in particular, see Tables 7 and 8 in there). 
	In this paper, we just highlight the crucial aspects taken under consideration, such as 
	\begin{itemize}
		\item[$i)$] the continuum fitting that allows to remove the long wavelength dependence induced by the quasar, 
		\item[$ii)$] the metal contamination, 
		\item[$iii)$] the thermal history of the intergalactic medium, 
		\item[$iv)$] the instrumental resolution and noise, 
		\item[$v)$] the impact of galactic feedback (Active Galactic Nuclei and Supernovae) and fluctuations of 
			the ultraviolet background, 
		\item[$vi)$] the impact of strong absorbers, 
		\item[$vii)$] the numerical modeling of small scales in hydrodynamical simulations. 	
	\end{itemize}
	As a final remark, Ref.\ \cite{Palanque-Delabrouille:2013gaa} compares two different methods of power spectrum extraction 
	and contrasts with the previous results by Ref.\ \cite{McDonald:2004eu}, finding very good agreement.
	For a comprehensive assessment of astrophysical effects on the flux power, we refer also to other analyses in 
	Refs.~\cite{Palanque-Delabrouille:2014jca,McDonald:2004xp,Viel:2005ha,%
	Palanque-Delabrouille:2015pga,Viel:2003fx,Viel:2004bf}, 
	while for uncertainties related to the method used in obtaining the limits, we refer to 
	Refs.~\cite{Palanque-Delabrouille:2014jca,Palanque-Delabrouille:2015pga}.

	For the purpose of the present paper, it is important to stress that these effects have been modeled by carefully 
	quantifying their impact in terms of flux power from state-of-the-art hydrodynamical simulations. 
	The amplitude and shape of such effect is then parameterized (template fitting) and marginalized over in the likelihood
	calculation. As a result, all the quoted relevant numbers already include the effect of the systematic uncertainties and 
	the dependence of the results upon the other cosmological parameters.  The relatively wide range of scales and 
	redshifts spanned by the observed one-dimensional flux power spectrum allows to effectively break the degeneracies 
	among astrophysical and cosmological parameters, providing tight constraints in terms of neutrino masses.

	We would like to comment on the widespread attitude to dismiss cosmological measurements, 
	invoking the possible existence of other sources of systematics, not yet identified, that could affect to some extent 
	the conclusion. This is in principle possible, as it is for any kind of measurement. 
	In some sense, it can be considered a super-conservative approach. However, we believe that the best thing that can be 
	done is identifying the possible relevant systematics and to quantify their impact, just as it was done 
	in Ref.~\cite{Palanque-Delabrouille:2013gaa}. This is especially important to continuously improve the reliability of 
	the present cosmological measurements, 
	also in view of their impact on the interpretation of \bb~in terms of massive neutrinos.
	
	From the frequentist interpretation (which is in excellent agreement with the bayesian results), 
	the authors of Ref.\ \cite{Palanque-Delabrouille:2014jca} compute a probability for $\Sigma$ that can be summarized in 
	a very a good approximation by:
	\begin{equation} \label{eq:chiCosm}
		\Delta \chi^2(\Sigma) = \frac{(\Sigma- 22\,\meV)^2}{(62\,\meV)^2}.
	\end{equation}
	Starting from the likelihood function $\mathcal{L}~\propto~\exp{-(\Delta \chi^2 /2)}$ with $\Delta \chi^2$ 
	as derived from Fig.~7 of Ref.~\cite{Palanque-Delabrouille:2014jca}, one can obtain the following limits:
	\begin{equation} \label{eq:clCosm}
		\begin{aligned}
			\Sigma < \hphantom{1}84\,\meV		&\qquad (1\sigma \,\mbox{C.\,L.}) \\
			\Sigma < 146\,\meV					&\qquad (2\sigma \,\mbox{C.\,L.}) \\
			\Sigma < 208\,\meV					&\qquad (3\sigma \,\mbox{C.\,L.})
		\end{aligned}
	\end{equation}
	which are very close to those predicted by the Gaussian $\Delta \chi^2$ of Eq.~(\ref{eq:chiCosm}). 
	
	It is worth noting that, even if this measurement is compatible with zero at less than $1\sigma$, the best fit value 
	is different from zero, as expected from the oscillation data and as evidenced by Eq.~(\ref{eq:chiCosm}).

	\begin{figure*}[tb]
		\centering
		\includegraphics[width=.95\columnwidth]{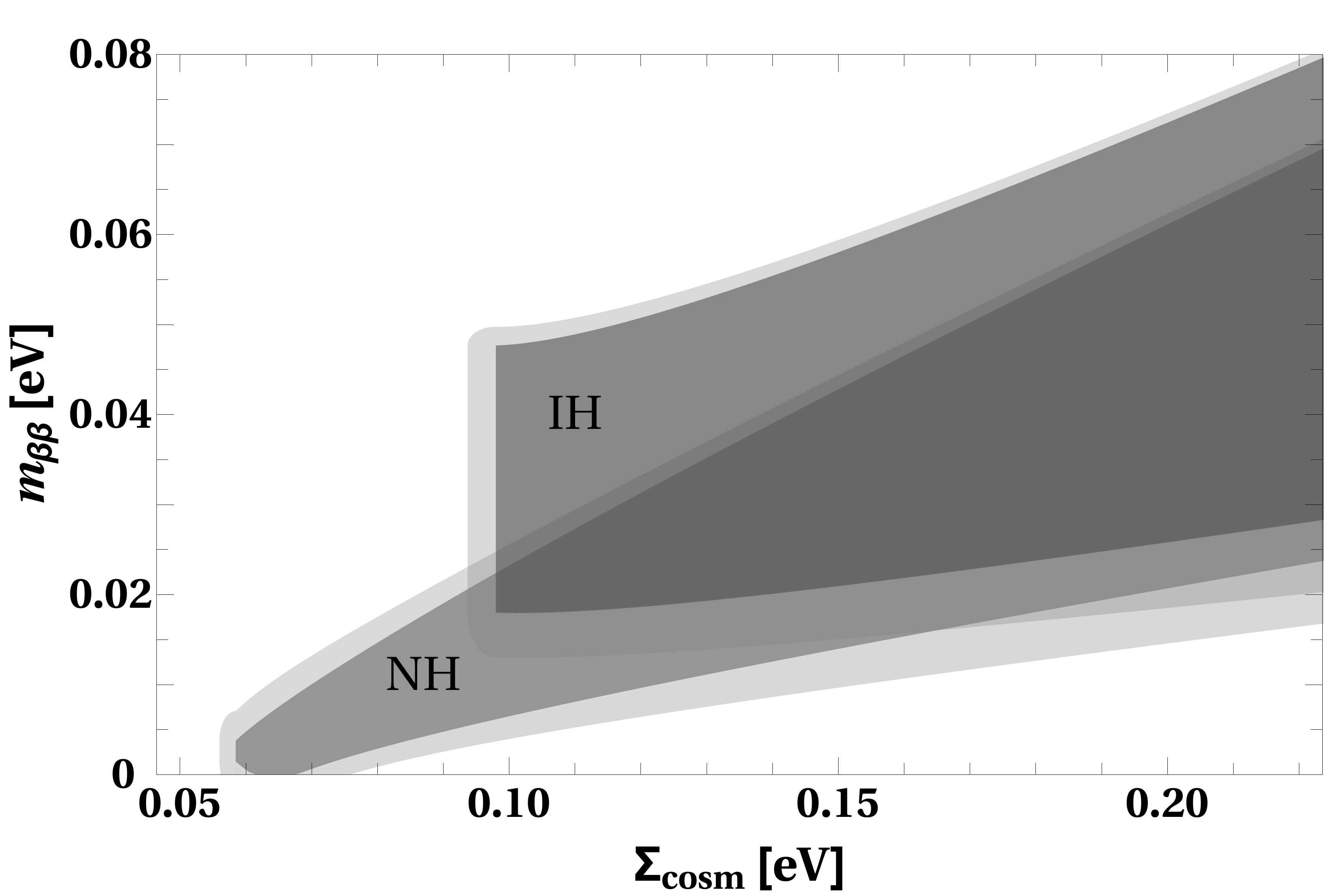}
		\quad
		\includegraphics[width=.95\columnwidth]{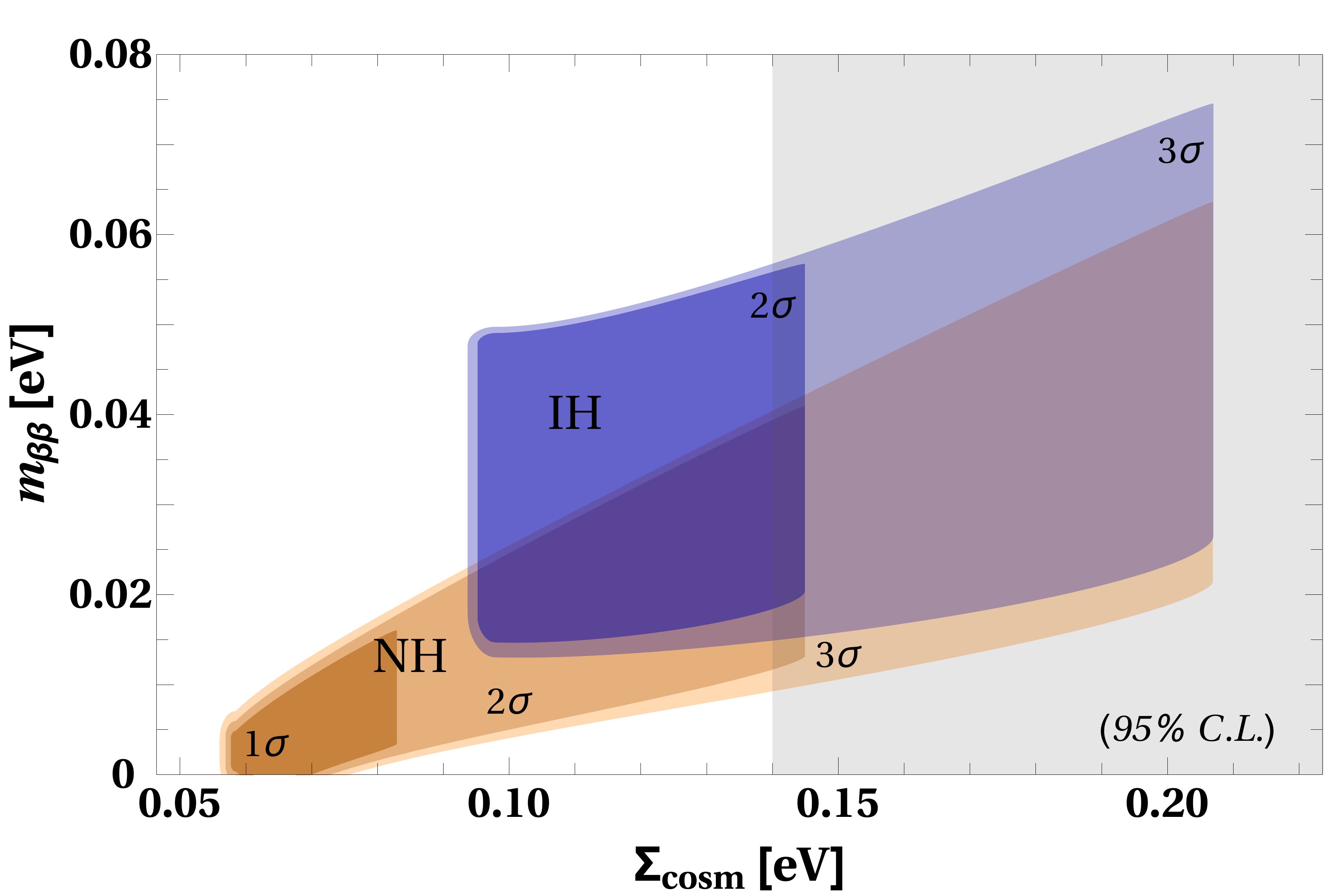}
		\caption{(\emph{Left}) Allowed regions for $\mbb$ as a function of $\Sigma$ with constraints given by 
			the oscillation parameters. The darker regions show the spread induced by Majorana phase variations, 
			while the light shaded areas correspond to the $3\sigma$ regions due to error propagation of the uncertainties on 
			the oscillation parameters.
			(\emph{Right}) Constraints from cosmological surveys are added to those from oscillations. 
			Different C.\,L.\ contours are shown for both hierarchies. Notice that the 1$\sigma$ region for the 
			\IH~case is not present, being the scenario disfavored at this confidence level. 
			The dashed band signifies the 95\% C.\,L.\ excluded region coming from Ref.~\cite{Palanque-Delabrouille:2014jca}.}
		\label{fig:blobs}
	\end{figure*}

	Furthermore, the (atmospheric) mass splitting $\Delta \equiv \sqrt{\Delta m^2} \simeq 49\,\meV$~\cite{Capozzi:2013csa} 
	becomes the dominant term of Eqs.\ \ref{eq:sigmaNH} and \ref{eq:sigmaIH} in the limit $m \to 0$. 
	Under this assumption, in the case of \NH~(\IH) $\Sigma$ reduces approximately to $\Delta$ $(2\Delta)$.
	This explains why this result favors the \NH~mass spectrum, as pointed out 
	in Ref.~\cite{Palanque-Delabrouille:2014jca} and as advocated in older theoretical works~\cite{Strumia:2005tc}.

	It is the first time that some data probing the absolute neutrino mass scale indicate a preference for one specific 
	mass hierarchy.
	Nonetheless, these results from cosmology have to be taken with due caution. In fact, the scientific literature contains 
	several authoritative claims for a non-zero value for $\Sigma$, 
	see e.\,g.\ Refs.\ \cite{Primack:1994pe,Allen:2003pta,Wyman:2013lza,Battye:2013xqa}. 
	In particular, it has been very recently suggested \cite{Wyman:2013lza,Battye:2013xqa} that a total non zero neutrino 
	mass around $0.3\,\eV$ could alleviate some tensions present between cluster number counts 
	(selected both in X-ray and by Sunyaev-Zeldovich effect) and weak lensing data.
	In some cases, a sterile neutrino particle with mass in a similar range is also
	advocated~\cite{Hamann:2013iba,Ade:2013zuv}.
	However, evidence for non-zero neutrino masses either in the active or sterile sectors seems to be claimed 
	in order to fix the significant tensions between different data sets (CMB and BAOs on one side and weak lensing, 
	cluster number counts and high values of the Hubble parameter on the other).  
	In fact, extending the model by including massive neutrinos does not improve much the agreement among
	different observables, since CMB and BAOs data do not support this extension~\cite{Leistedt:2014sia}. 
	This could suggest that systematic errors are not fully under control.
	More precise measurements from cosmological surveys are expected in the near future 
	(among the others, DESI%
	\footnote{\href{http://desi.lbl.gov}{http://desi.lbl.gov}} 
	and the Euclid satellite%
	\footnote{\href{http://www.euclid-ec.org}{http://www.euclid-ec.org}}) 
	and they will probably allow more accurate statements on neutrino masses.
	
\section{Implication for the \bb~search}

	The close connection between the results on neutrino mass obtained in the laboratory and those coming from cosmology 
	was outlined long ago~\cite{Zeldovich:1981wf}. 
	In order to discuss the mass parameter $\mbb$ which is relevant for the \bb, it is useful to adopt the 
	representation introduced in Ref.~\cite{Fogli:2004as}, 
	where $\mbb$ is expressed as a function of $\Sigma$. The result is shown in the left panel of Fig.~\ref{fig:blobs}.
	The procedure to obtain this plot is quite straightforward:
	\begin{itemize}
		\item[-] the starting points are the expressions for the maximum and the minimum values of $\mbb$ as a function of 
			the lightest neutrino mass $m$, the oscillation parameters and the unknown Majorana phases. These last are left 
			free to vary between their extremal values, thus providing an interval of values for $\mbb$ for a given set 
			of $m$ and oscillation parameters (see e.\,g.\ Refs.~\cite{Vissani:1999tu,DellOro:2014yca} for a more 
			detailed explanation).
		\item[-] for both the mass hierarchies, one has to solve the quartic equation that gives $m$ as a function of 
			$\Sigma$, $\delta m^2$ and $\Delta m^2$. 
			This solution is replaced in the previous expressions of $\mbb$, which in turn become a function of $\Sigma$ and 
			of the oscillations parameters.
		\item[-] the residual uncertainties on the oscillation parameters must be propagated on the maximum and minimum values 
			of $\mbb$ (the procedure is described e.\,g.\ in Ref.\cite{DellOro:2014yca}).
			This operation slightly widens the allowed regions for $\mbb$.
		\item[-] the minimum value of $\Sigma$, namely $\Sigma(m=0, \delta m^2,\Delta m^2)$, is different from zero for both 
			the mass hierarchies and its value is subject to the residual uncertainties on $\delta m^2$ and $\Delta m^2$.
			These can be easily propagated as follows:
			\begin{equation}
				\delta \Sigma = \sqrt{	
						\left(\frac{\partial\,\Sigma \hspace{11pt}}{\partial\,\delta m^2}\,\sigma(\delta m^2) \right)^2 + 
						\left(\frac{\partial\,\Sigma \hspace{11pt}}{\partial\,\Delta m^2}\,\sigma(\Delta m^2) \right )^2}\,.
			\end{equation}
			This explains the widening of the shaded areas in the leftmost part of the allowed regions in the left panel 
			of Fig.~\ref{fig:blobs}.
	\end{itemize}
	
	The new cosmological constraints on $\Sigma$ from Ref.~\cite{Palanque-Delabrouille:2014jca} can be now included at 
	any desired confidence level by considering the following inequality:
	\begin{equation}
		\frac{(y-\mbb(\Sigma))^2}{ (n\,\sigma[ \mbb(\Sigma) ])^2}+\frac{(\Sigma-\Sigma(0))^2}{(\Sigma_n-\Sigma(0)) ^2}<1
		\label{eq:chiblob}
	\end{equation}
	where $\mbb(\Sigma)$ is the Majorana Effective Mass as a function of $\Sigma$ and
	$\sigma[\mbb(\Sigma)]$ is the 1$\sigma$ associated error, computed as discussed in Ref.~\cite{DellOro:2014yca}.
	$\Sigma_n$ is the limit on $\Sigma$ derived from Eq.~(\ref{eq:chiCosm}) for the C.\,L.\ $n=1,2,3,\dots$
	By solving Eq.~(\ref{eq:chiblob}) for $y$, it is thus possible to get the allowed contour for $\mbb$ considering both the 
	constraints from oscillations and from cosmology. In particular, the Majorana phases are taken into account by 
	computing $y$ along the two extremes of $\mbb(\Sigma)$, namely $\mbb^{max}(\Sigma)$ and $\mbb^{min}(\Sigma)$, 
	and then connecting the two contours. The resulting plot is shown in the right panel of Fig.\ \ref{fig:blobs}.
	
	The most evident feature of Fig.\ \ref{fig:blobs} is the clear difference in terms of expectations for both $\mbb$ 
	and $\Sigma$ in the two hierarchy cases. 
	The relevant oscillation parameters (mixing angles and mass splittings) are well known and they induce only 
	minor uncertainties on the expected value of $\mbb$. These uncertainties widen the allowed contours in the upper, 
	lower and left sides of the picture.
	The boundaries in the rightmost regions are due to the new information from cosmology and are cut at various 
	confidence levels. It is notable that at 1$\sigma$, due to the exclusion of the 
	\IH, the set of plausible values of $\mbb$ is highly restricted.
	Table \ref{tab:mbb_bounds} summarizes the new allowed values for $\mbb$.


	\begin{table}[tb]
		\centering
		\begin{tabular}{lrrr}
			\hline \\[-12pt] \hline \\[-10pt]
			Mass spectrum								&\multicolumn{3}{c}{$\mbb$ max [meV] (C.\,L.\ on $\Sigma$)}		\\
			\cline{2-4}
											&$1\sigma$	\hspace{30pt}		&$2\sigma$		&$3\sigma$	\\
			\hline
			\NH							&16 \hspace{30pt}					&41				&64			\\
			\IH							&- \hspace{30pt}					&57				&75			\\
			\hline \\[-12pt] \hline 
		\end{tabular}
		\caption{Maximum values for $\mbb$ once the new constraints on $\Sigma$ 
			from Ref.~\cite{Palanque-Delabrouille:2014jca} are added. The $1\sigma$ C.\,L. maximum value on $\mbb$ for the \IH~is
			not reported because the scenario is excluded in this case.}
		\label{tab:mbb_bounds}
	\end{table}

	The next generation of \bb~experiments is expected to probe the upper values of the predicted \IH~region and reach 
	a sensitivity for $\mbb$ of about $70\,\meV$~\cite{DellOro:2014yca}. This assumes the absence of the 
	quenching of the axial coupling constant~\cite{Faessler:2007hu,Barea:2013bz,Robertson:2013cy}, 
	which would imply even longer lifetimes and conversely would worsen the sensitivity to $\mbb$ 
	significantly~\cite{DellOro:2014yca}.

\begin{figure}[b]
		\centering
		\includegraphics[width=\columnwidth]{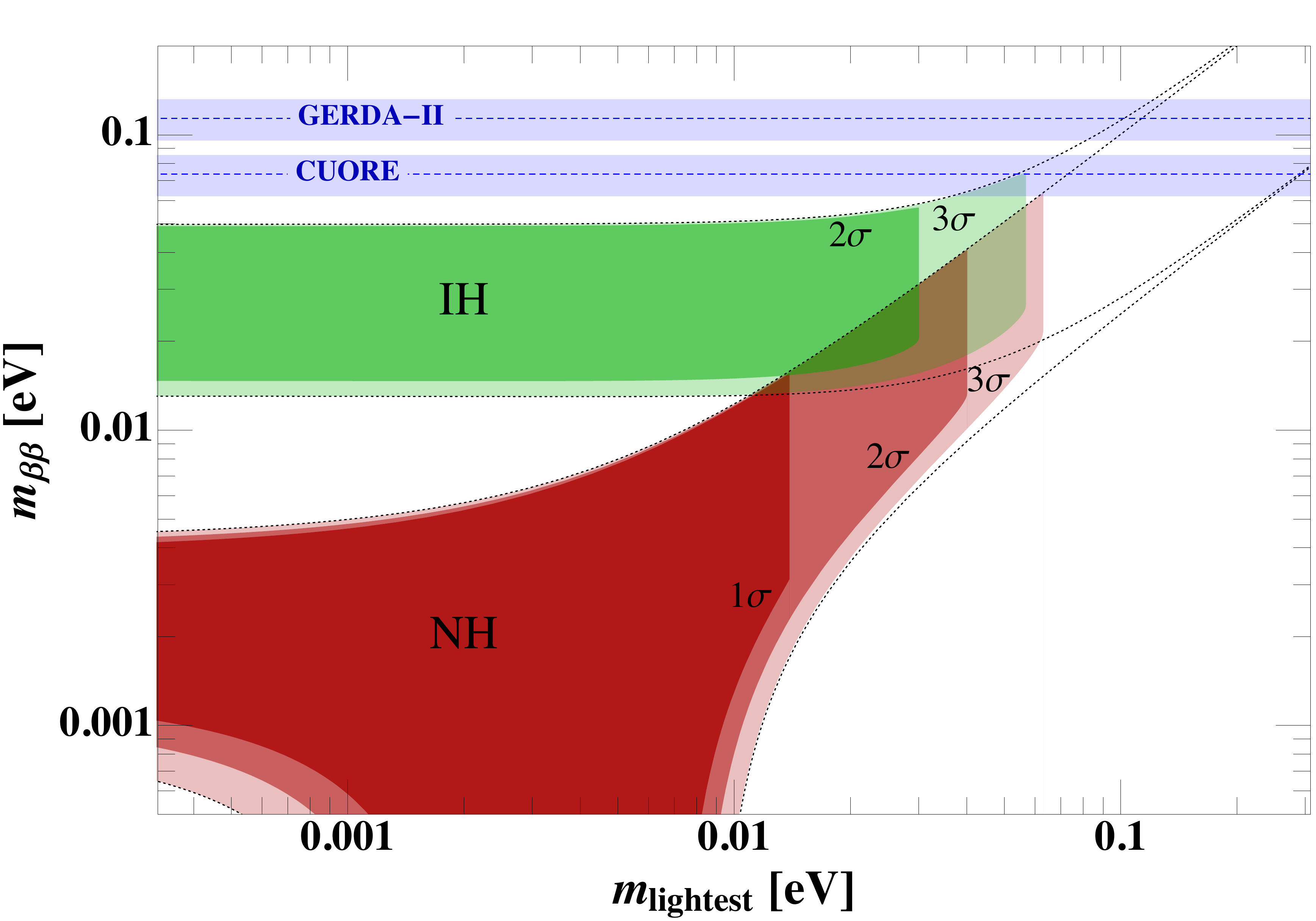}
		\caption{Constraints from cosmological surveys are added to those from oscillations in the representation
			$\mbb$ as a function of the lightest neutrino mass. 
			The dotted contours represent the $3\sigma$ regions allowed considering oscillations only.
			The shaded areas show the effect of the inclusion of cosmological constraints at different C.\,L.\,. 
			The horizontal bands correspond to the expected sensitivity for future experiments, 
			according to Ref.\ \cite{DellOro:2014yca}.}
		\label{fig:VS}
	\end{figure}

	The impact of the new constraints on $\Sigma$ appears even more evident by plotting $\mbb$ as a function of the mass of the 
	lightest neutrino~\cite{Vissani:1999tu}.
	In this case, Eq.~(\ref{eq:chiblob}) becomes:
	\begin{equation}
		\frac{(y-\mbb(m))^2}{ (n\,\sigma[ \mbb(m) ])^2}+\frac{m^2}{m(\Sigma_n) ^2}<1.
		\label{eq:chiblobVS}
	\end{equation}
	Here, $\mbb(m)$ is $\mbb$ expressed as a function of $m$, $\sigma[\mbb(m)]$ is the 1$\sigma$ associated error, computed 
	as discussed in Ref.\ \cite{DellOro:2014yca}, and $m(\Sigma_n)$ is the value of $m$ calculated for a given $\Sigma_n$.
	In Fig.~\ref{fig:VS} the expected sensitivities for two examples of next generation
	\bb~experiments (CUORE~\cite{Artusa:2014lgv} and GERDA-II~\cite{Brugnera:2013xma}) are presented. 
	The dashed contours indicate the $3\sigma$ regions allowed by oscillations. 
	The shaded areas are the new allowed regions once the cosmological constraints are added.
	The plot globally shows that the next generation of experiments will have small possibilities of 
	detecting a signal of \bb~due to light Majorana neutrino exchange.
	Therefore, if the new results from cosmology are confirmed or improved, ton or even multi-ton scale detectors 
	will be needed~\cite{DellOro:2014yca}.

	On the other hand, a \bb~signal in the near future could either disprove some assumptions of the present cosmological 
	models, or suggest that a different mechanism other than the light neutrino exchange mediates the transition.
	New experiments are interested in testing the latter possibility by probing scenarios beyond the 
	Standard Model~\cite{Cirigliano:2004tc,Tello:2010am,Alekhin:2015oba}.  

\section{Conclusion}

	A cautious attitude in dealing with the results from cosmological surveys is highly advisable. However, the newest 
	analysis~\cite{Palanque-Delabrouille:2015pga} confirms and strengthens the cosmological indications, and it is likely 
	that we will have soon other substantial progress. 
	Moreover, the present theoretical understanding of neutrino masses does not contradict these cosmological indications.
	These considerations emphasise the importance of exploring the issue of mass hierarchy in laboratory experiments and 
	with cosmological surveys. 
	
	From the point of view of \bb, these results show that ton or multi-ton scale detectors will 
	be needed in order to probe the range of $\mbb$ now allowed by cosmology.
	Nevertheless, if next generation experiments see a signal, it will likely be a \bb~signal of new physics 
	different from the light Majorana neutrino exchange.


	\acknowledgments
		We gratefully thank Prof.\ F.\ Iachello for his suggestions and illuminating discussions on 
		neutrinoless double beta decay.
		We warmly thank K.\ Randle for carefully checking the text.
		M.\ V.\ is supported by the ERC-StG cosmoIGM and by INFN-PD51 IS Indark.

	\smallskip
	\paragraph*{Note added:}


	After this work was completed, new completely independent analysis with results
	fully consistent with the one discussed in Ref.~\cite{Palanque-Delabrouille:2014jca} were presented.
	Indeed, in Refs.~\cite{Cuesta:2015iho} and \cite{Zhang:2015uhk}, the limits $\Sigma< 0.11$ and $\Sigma< 0.113\,\eV$ at 
	$95\%$\,C.\,L., respectevely, were derived by combining data from BAOs, CMB and galaxy clustering. 
	This adds confidence to our hypotheses and inferences.

\bibliography{ref}
	
\end{document}